\def\figsize{9.5cm}

\def\rn{}
\def\nn#1 #2{#2. #1}                
\def\nnn#1 #2 #3{#2. #3. #1}            
\def\nnnn#1 #2 #3 #4{#2. #3. #4 #1}     
\def\nnnnn#1 #2 #3 #4 #5{#2. #3. #4 #5. #1} 

\def\rf#1;#2;#3;#4;#5 {{\frenchspacing\par\rn#1, #3 {\bf #4}, #5 (#2). \par}}
\def\rrf#1;#2;#3;#4;#5 {{\frenchspacing\rn#1, #3 {\bf #4}, #5 (#2);~}}
\def\rrrf#1;#2;#3;#4;#5 {{\frenchspacing\rn#1, #3 {\bf #4}, #5 (#2).}}
\def\rg#1;#2;#3;#4;#5;#6 {{\frenchspacing\par\rn#1, #3 {\bf #4}, #5 (#2). \par}}
\def\rfbook#1;#2;#3;#4;#5 {{\frenchspacing\par\rn#1, {\it #3} (#5, #4, #2).\par}}
\def\rfprep#1;#2;#3 {{\par\frenchspacing\rn#1, #3 (#2).\par}}
\def\rrfprep#1;#2;#3 {{\frenchspacing\rn#1, #3 (#2);~}}
\def\rrrfprep#1;#2;#3 {{\frenchspacing\rn#1, #3 (#2).}}
\def\rfproc#1;#2;#3;#4;#5;#6 {{\frenchspacing\par\rn#1 #2, in {\it #3}, ed. #4 (#5: #6)\par}}
\def\rfprocp#1;#2;#3;#4;#5;#6;#7 {{\frenchspacing\par\rn#1 #2, in {\it #3}, ed. #4 (#5: #6), p#7\par}}

\def\rg#1;#2;#3;#4;#5;#6 {\par\rn#1 #2, {\it #3}, {\bf #4}, #5 (``#6'') \par}
\def\rf#1;#2;#3;#4;#5 {\par\rn#1, {\it #3}, {\bf #4}, #5 (#2)\par}
\def\rfbook#1;#2;#3;#4;#5 {{\frenchspacing\par\rn#1, {\it #3} (#4: #5, #2)\par}}
\def\rfproc#1;#2;#3;#4;#5;#6 {{\frenchspacing\par\rn#1 #2, in {\it #3}, ed. #4 (#5: #6)\par}}
\def\rfprocp#1;#2;#3;#4;#5;#6;#7 {{\frenchspacing\par\rn#1 #2, in {\it #3}, ed. #4 (#5: #6), p#7\par}}
\def\rfprep#1;#2;#3  {{\par\rn#1, #3, #2\par}}
\def\rfprepp#1;#2;#3 {{\par\rn#1 #2, #3\par}}

\def\etal{{\frenchspacing\it et al.}}

\def\beq#1{\begin{equation}\label{#1}}
\def\eeq{\end{equation}}
\def\beqa#1{\begin{eqnarray}\label{#1}}
\def\eeqa{\end{eqnarray}}

\def\spose#1{\hbox to 0pt{#1\hss}}
\def\simlt{\mathrel{\spose{\lower 3pt\hbox{$\mathchar"218$}}
     \raise 2.0pt\hbox{$\mathchar"13C$}}}
\def\simgt{\mathrel{\spose{\lower 3pt\hbox{$\mathchar"218$}}
     \raise 2.0pt\hbox{$\mathchar"13E$}}}
\def\simpropto{\mathrel{\spose{\lower 3pt\hbox{$\mathchar"218$}}
     \raise 2.0pt\hbox{$\propto$}}}

\def\ed{\end{document}}

\def\f{X}

\def\beq#1{\begin{equation}\label{#1}}
\def\eeq{\end{equation}}
\def\beqa#1{\begin{eqnarray}\label{#1}}
\def\eeqa{\end{eqnarray}}

\documentclass[twocolumn,amsmath,nofootinbib]{revtex4} 
\usepackage{graphicx}
\usepackage{url}
\begin{document}
\input{epsf.sty}

\def\affilmrk#1{$^{#1}$}
\def\affilmk#1#2{$^{#1}$#2;}

\title{Dark Energy Constraints from the Cosmic Age and Supernova}

\author{Bo Feng$^1$, Xiulian Wang$^{2}$, and Xinmin Zhang$^{1}$ }
\affiliation{$^{1}$Institute of High Energy Physics, Chinese
Academy of Science, P.O. Box 918-4, Beijing 100039, P. R. China}
\affiliation{$^{2}$Institute of Theoretical Physics, Chinese
Academy of Sciences, Beijing 100080, P. R. China.}

\begin{abstract}
Using the low limit of cosmic ages from globular cluster and the
white dwarfs: $t_0 > 12$Gyr, together with recent new high
redshift supernova observations from the HST/GOODS program and
previous supernova data, we give a considerable estimation of the
equation of state for dark energy, with uniform priors as weak as
$0.2<\Omega_m<0.4$ or $0.1<\Omega_m h^2<0.16$. We find cosmic age
limit plays a significant role in lowering the upper bound on the
variation amplitude of dark energy equation of state. We propose
in this paper a new scenario of dark energy dubbed Quintom, which
gives rise to the equation of state larger than $-1$ in the past
and less than $-1$ today, satisfying current observations. In
addition we've also considered the implications of recent X-ray
gas mass fraction data on dark energy, which favors a negative
running of the equation of state.

\end{abstract}


\maketitle


\setcounter{footnote}{0}


Age limits of our universe are among the earliest motivations for
the existence of the mysterious dark energy. Namely, observations
of the earliest galaxies could set a low limit on the age of the
universe. In 1998, two groups \cite{Riess98,Perl99} independently
showed the accelerating expansion of our universe basing on Type
Ia Supernova (SNe Ia) observations of the redshift-distance
relations. The recently released first year WMAP data
\cite{Spergel03} support strongly the concordance model with dark
energy taking part of $\sim 2/3$. The most recent discovery of 16
SNe Ia \cite{Riess04} with the Hubble Space Telescope during the
GOODS ACS Treasury survey, together with former SNe Ia data alone
could provide a strong hint for the existence of dark energy.
Riess {\it et al.}\cite{Riess04} provided evidence at $>99 \%$ for
the existence of a transition from deceleration to acceleration
using supernova data alone.

Despite our current theoretical ambiguity for the nature of dark
energy, the prosperous observational data ({\it e.g.} supernova,
CMB and large scale structure data and so on ) have opened a
robust window for testing the recent and even early behavior of
dark energy using some simple parameterization for its equation of
state ({\it e.g.}, Ref. \cite{linder} ) or even reconstruction of
its recent density \cite{sahni00,dd03,wangy04}. Both recent WMAP
fit and more recent fit by Riess {\it et al.} find the behavior of
dark energy is to great extent in consistency with a cosmological
constant. In particular when the equation of state is not
restricted to be a constant, the fit to observational data
improves dramatically \cite{sahni,cooray,wangy03,gong}. Huterer
and Cooray \cite{cooray} produced uncorrelated and nearly
model-independent band power estimates (basing on the principal
component analysis\cite{Huterer}) of the equation of state of dark
energy and its density as a function of redshift, by fitting to
the recent SNe Ia data they found marginal (2-$\sigma$) evidence
for $W(z) < -1$ at $z < 0.2$, which is consistent with other
results in the
literature\cite{sahni,cooray,wangy04,wangy03,sahni03,Nesseris,CP03,zhu}.

The recent fit to first year WMAP and other CMB data, SDSS and 172
SNe Ia data \cite{Tonry03} by Tegmark et al \cite{0310723}
provided the most complete and up-to-date fit. Although SNe Ia
data accumulated more after that, Ref. \cite{0310723} should still
be a very profitable benchmark for current fit of the observables.
However, when considering the behavior of dark energy alone, one
has to do more since Ref.\cite{0310723} only dealt with constant
equation of state before the recent release of 16 more SNe Ia data
by Ref.\cite{Riess04}. In fact a complete fit to full
observational data still remains impossible provided one wants to
reconstruct the full behavior of dark energy, despite the using of
the most efficient Markov Chain Monte Carlo(MCMC) method. Under
such circumstance Wang {\it et al.} and Riess {\it et al.} fitted
dark energy to SNe Ia data, 2df\cite{2df} linear growth factor and
a parameter related (up to a constant) to the angular size
distance to the last scattering surface. In fact even the angular
size distance is model dependent, as can be seen from
Ref.\cite{Spergel03} it differs for the six-parameter vanilla
model and when an additional parameter $\alpha$ (running of the
spectral index) is added. Regarding the constraint from the cosmic
age, Krauss\cite{Krauss04} used the WMAP fitted value for seven
parameters: $t_0= 13.7 \pm 0.2$Gyr, together with $1\sigma$ HST
\cite{HST} bound and assuming some specific relation between
$\Omega_m$ and $h$, he got a lower bound on constant equation of
state: $W>-1.22$, which is in strikingly agreement with WMAP
result. Generally speaking, age limit can give an upper limit
rather than lower limit, as shown by Cepa\cite{0403616}.

The low limit to the cosmic age can be directly obtained from
dating the oldest stellar populations. Globular clusters (GC) in
the Milky Way are excellent laboratory for constraining cosmic
ages. Carretta {\it et al.} \cite{carretta} gave the best estimate
for the age of GCs to be Age=$12.9\pm 2.9$Gyr at $95\%$ level. The
limit for age of GCs is around 11-16 Gyr\cite{Spergel03}. White
dwarf dating provides a good approach to the main sequence
turn-off. Richer {\it et al.}\cite{Richer} and Hansen {\it et
al.}\cite{Hansen} found an age of $12.7\pm 0.7$Gyr at $2\sigma$
level using the white dwarf cooling sequence method. For a full
review of cosmic age limit see Ref.\cite{Spergel03}. The low limit
to cosmic age serves as the `anti-smoking gun' in excluding models
which lead to shorter age. In this paper we use $t_0>12.0$Gyr as
the bound on cosmic age. Other constraints we use are only
uniformly in range $0.2<\Omega_m<0.4$ {\bf or} $0.1<\Omega_m
h^2<0.16$. As can be seen from Ref.\cite{0310723}, such constraint
is much looser than the six parameters + $ W$ set and comparable
to the nine-parameter-vary set \footnote{Our prior on $\Omega_m$
is also consistent with the median statistics study on mass
density by Chen and Ratra\cite{CR03}, for more investigations on
the effects of priors see  Refs.\cite{CMMS04,wangy04,0405218}.}.
 We {\bf assume} this
constraint to be reasonable with one additional parameter added
below: the variation of equation of state. We find for the linear
parameterization of W age constraint can shrink the upper bound on
$W'$ from $\sim$15 to $\sim$5 when using SNe Ia alone and from
$\sim$5 to $\sim$2 when considering above priors on $\Omega_m$ or
$\Omega_m h^2$. While for the model introduced by
Linde\cite{linder}, the upper bound on $W_a$ can shrink from
$\sim$10 to $\sim$5 when with priors on $\Omega_m$ or $\Omega_m
h^2$.

\begin{figure}[htbp]
\begin{center}
\includegraphics[scale=0.45]{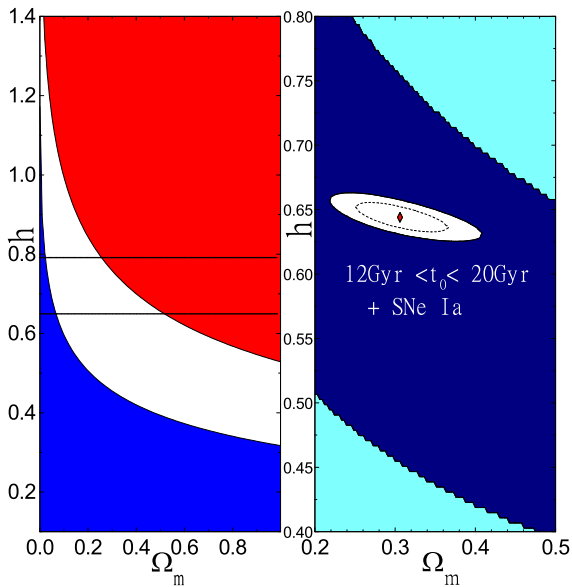}
\caption{ Age and SNe Ia constraints on $\Lambda$CDM cosmology.
Left panel: The red area is excluded by $t_0>12$Gyr and the blue
area is excluded by the assumption for $t_0<20$Gyr for
$\Lambda$CDM model. The area between the two solid lines is
allowed by $1\sigma$ HST limit. Right panel: $2\sigma$ SNe Ia
limit on $\Lambda$CDM model. The dashed line corresponds to the
$1\sigma$ limit and the dot inside denotes the best fit value. The
navy area is allowed by age constraint  12Gyr $<$ $t_0$ $<$ 20
Gyr. \label{fig:fig1}}
\end{center}
\end{figure}

The cosmic age can be written as
\begin{equation}
t_0 = H_0^{-1}\int^{\infty} _{0} \frac{dz}{(1+z) E(z)}~,
\end{equation}
where \cite{wangy04}
\begin{equation}
E(z) \equiv \left[\Omega_m (1+z)^3 +(1-\Omega_m)
\f(z)\right]^{1/2}
\end{equation}
and $X(z)\equiv\rho_X(z)/\rho_X(0)$.

 Firstly we delineate the effect
of age limit in $\Lambda$CDM cosmology. In the full paper we
assume a flat space, {\it i.e.} $\Omega_k=0$. In the left panel of
Fig. 1 we vary $\Omega_m$ from 0 to 1 and the Hubble parameter $h$
from 0 to 1.4. The red area is excluded by $t_0> 12.0$Gyr, the
area between the two black solid lines is given by the $1\sigma$
HST limit. If we conservatively assume in $\Lambda$CDM cosmology
the cosmic age is no more than 20 Gyr, the area with blue color
will be excluded. It can give a rough estimate for current
fraction of matter in the universe, although much looser than SNe
Ia constraint as shown in the right panel. The supernova data we
use is the "gold" set of 157 SNe Ia published by Riess {\etal} in
\cite{Riess04}. In the below we constrain the Hubble parameter to
be uniformly in $3\sigma$ HST region: $0.51<h<0.93$.

In the detailed discussions below we consider two type of
parameterizations for the equation of the state of the dark
energy,
\begin{equation}
W(z)=W+W'z,
\end{equation}
(taken as Model A) so that
\begin{equation}
\f=a^{-3(1+W-W')} e^{3W'(a^{-1}-1)}~,
\end{equation}
where $a$ is the cosmic scale factor. The other form was proposed
by Refs. \cite{DP,linder}(taken as Model B):
\begin{equation}
W(z)=W_1+W_a z/(1+z)~,
\end{equation}
which leads to
\begin{equation}
\f=a^{-3(1+W_1+W_a)} e^{3W_a(a-1)}~.
\end{equation}
Both models(as well as other models such as firstly proposed in
\cite{sahni03}) make good approximations to probe the behavior of
dark energy around the present epoch, while the former model leads
to poor parameterization at very large redshift. But as argued by
Riess {\etal} \cite{Riess04} this is acceptable for showing the
late behavior of dark energy.

\begin{figure}
\centerline{\epsfxsize=\figsize\epsffile{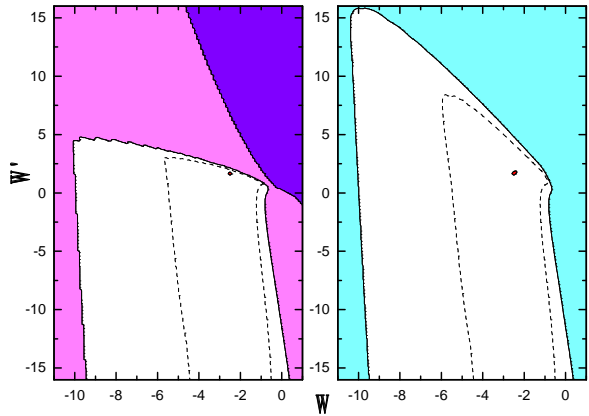}}
\caption[1]{\label{SausageFig}\footnotesize%
Right panel: $2\sigma$ SNe Ia limit alone on Model A dark energy.
Left panel: $2\sigma$ SNe Ia limit and age limit ($t_0>12$Gyr) on
Model A dark energy. The dots inside the two panels show the best
fit parameters.}
\end{figure}

SNe Ia data alone proves to be a weak constraint on above models,
as shown in the right panel of Fig.2 and also in
Ref.\cite{wangy04} where the authors used flux averaging
method\cite{wangyflux,wangy03}. In the left panel of Fig.2, the
region on up right corner is fully excluded by the age
limit($t_0>12$Gyr). The role of age limit can be easily seen from
Eqs.(1-6). In Model A larger $W'$ leads to larger X in early
epochs, hence corresponds to smaller ages which can be directly
constrained. Similar case works for Model B. Age constraint still
works when adding the prior on $\Omega_m$ or $\Omega_m h^2$. In
Figs.3-4 we show the corresponding effects when adding different
priors for SNe Ia on right panels and show the role of age
correspondingly in the left panels. The dashed lines are the
$1\sigma$ regions and the small dots denote the best fit
parameters. Up right regions of the left panels are excluded by
the cosmic age limit. We can see the age limit reduces
significantly the upper regions and consequently changes the best
fit values of the model parameters. One can also find from Fig.2
and Fig.3 the role of priors in deriving $w$, although the effect
of age limit does not change, the priors on $\Omega_m$ has
shrinked the allowed parameter space on $w$. Generically the
results on $w$ also depends on different parameterizations, as
shown in our paper and also in the literature($e.g.$ Ref.
\cite{sahni03}). However this would not change the picture that
cosmic age can put an additional constraint on the equation of
state $w$.

\begin{figure}
\centerline{\epsfxsize=\figsize\epsffile{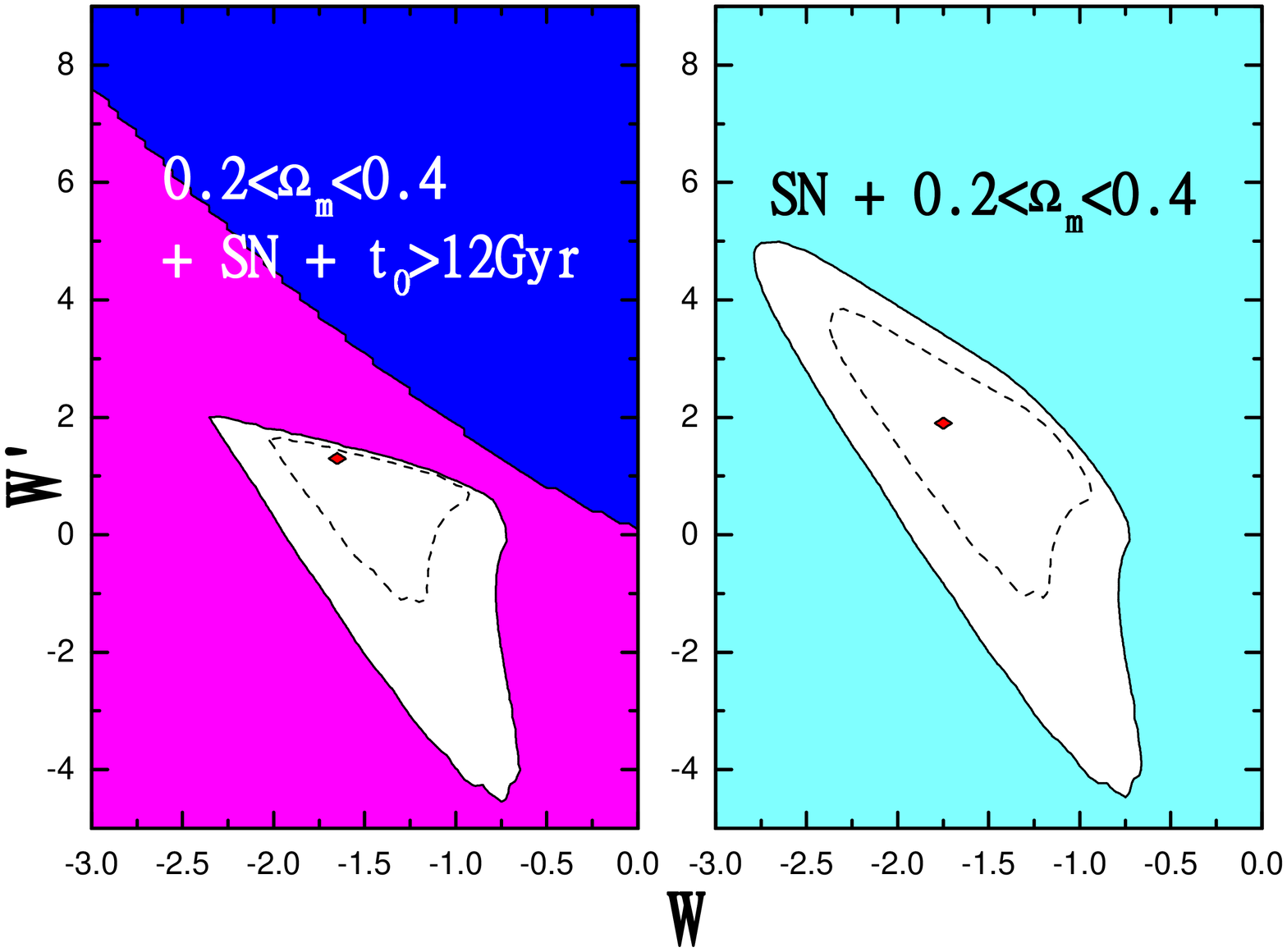}}
\centerline{\epsfxsize=\figsize\epsffile{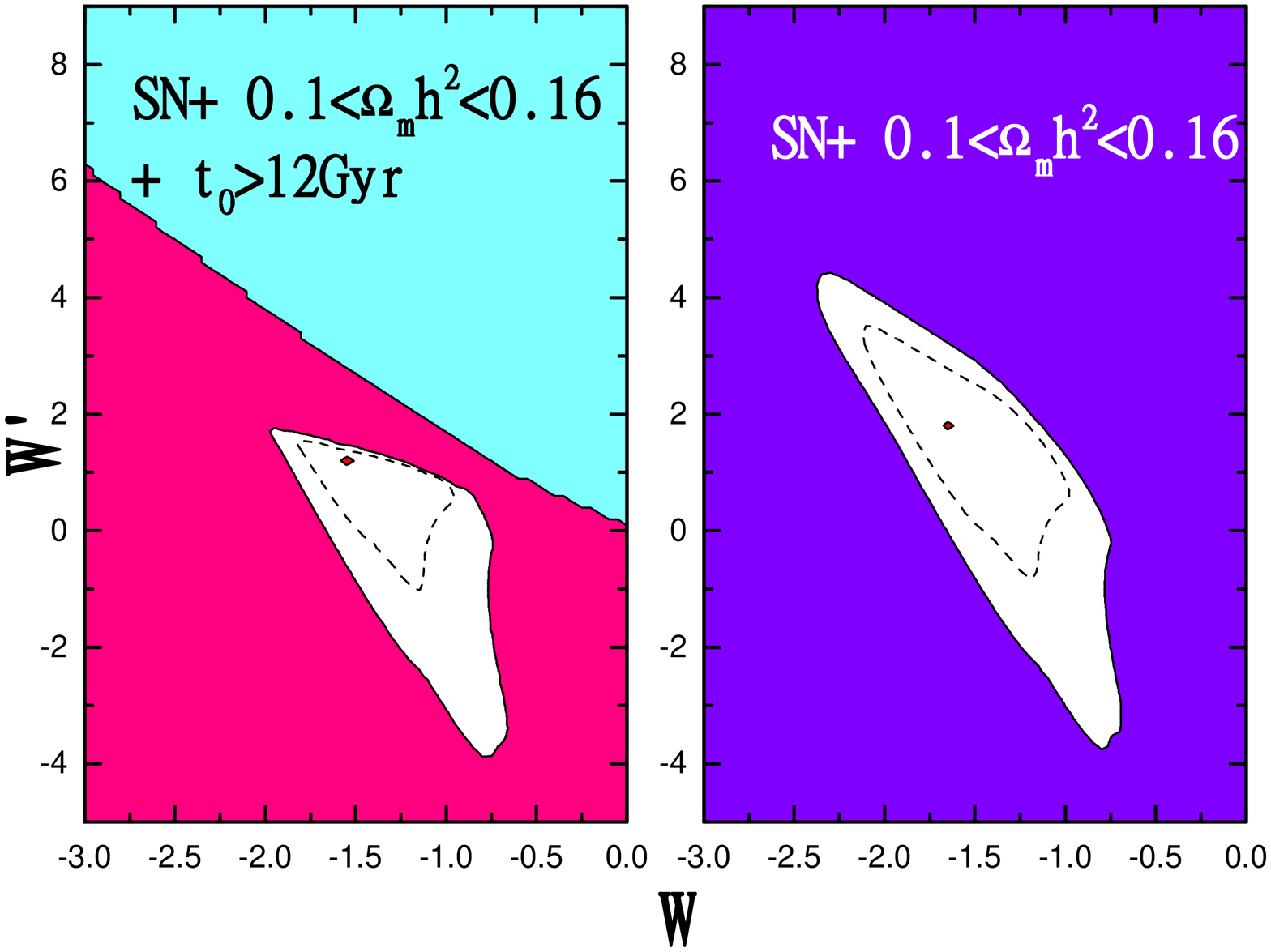}}
\caption[1]{\label{SausageFig}\footnotesize%
Age and SNe limits on Model A with different priors as noted
inside. The dots inside the 1$\sigma$ dashed lines denote the best
fit parameters. }
\end{figure}

\begin{figure}
\centerline{\epsfxsize=\figsize\epsffile{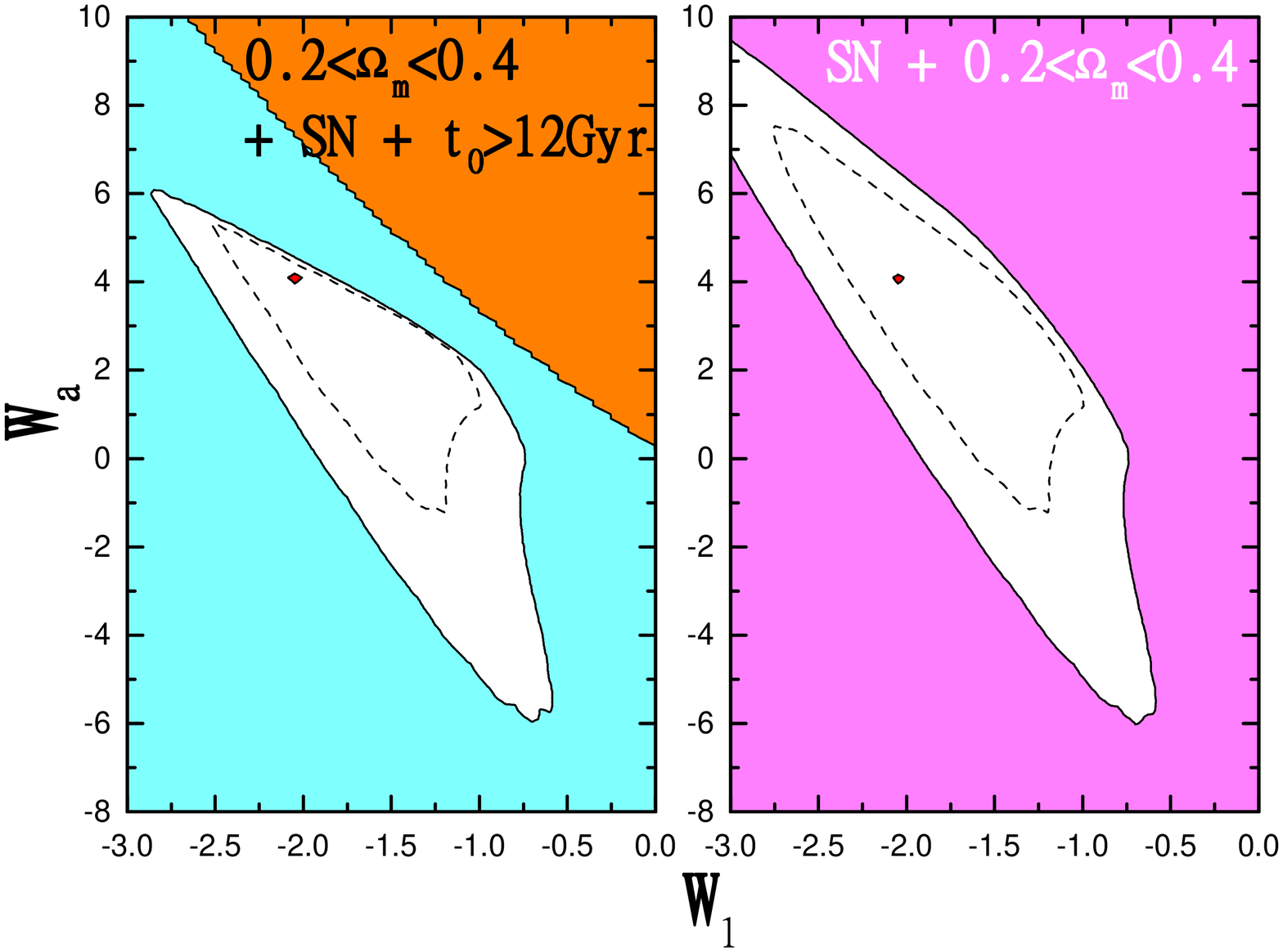}}
\centerline{\epsfxsize=\figsize\epsffile{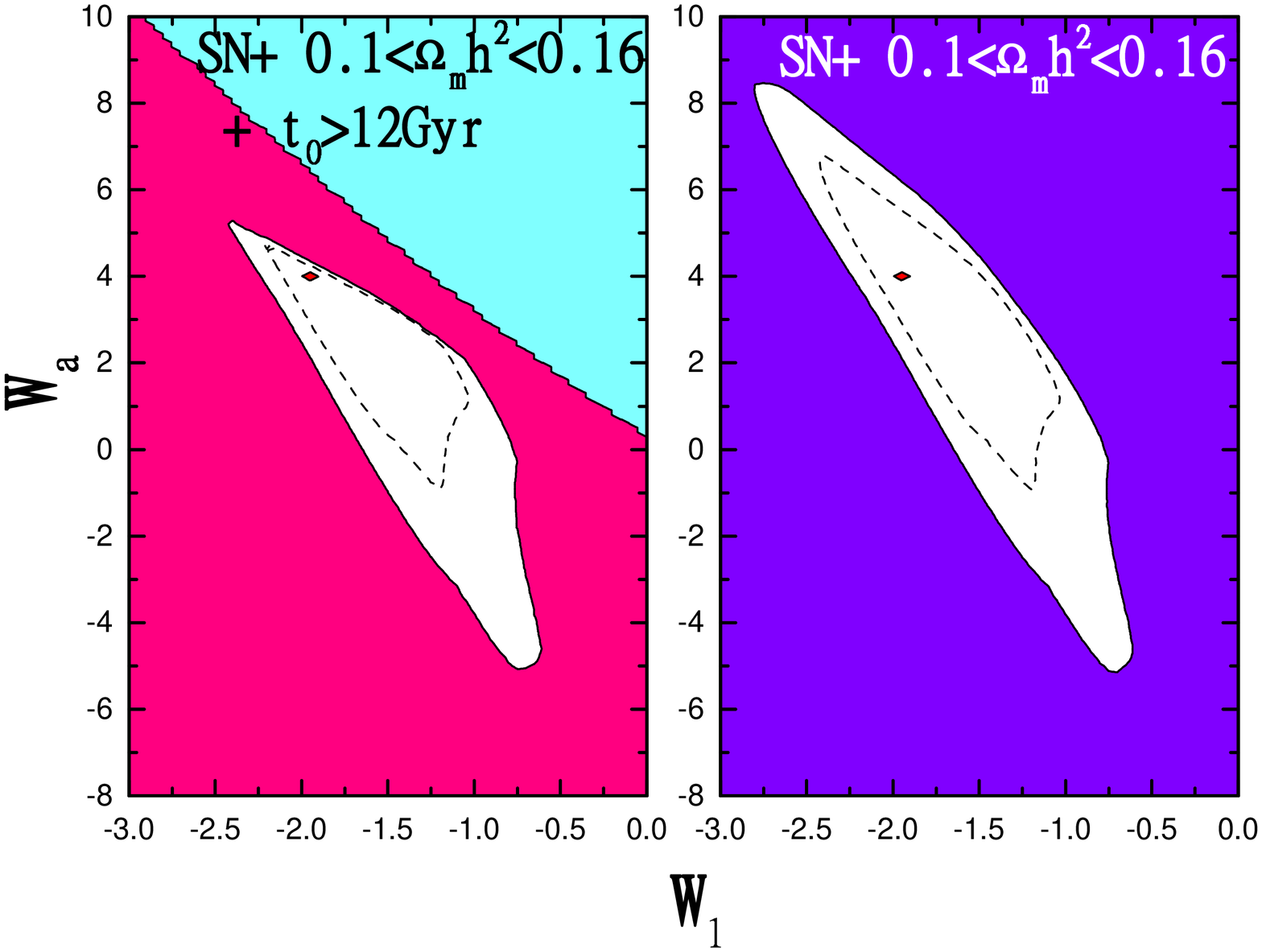}}
\caption[1]{\label{SausageFig}\footnotesize%
The same as Fig.3 for Model B.}
\end{figure}

As mentioned above, the present data seem to favor an evolving
dark energy with the equation of state being below $-1$ around
present epoch evolved from $W>-1$ in the past. This can also be
seen from the best fit parameters of our Figs. 2-5. If this result
is confirmed in the future, it has important implications for the
theory of dark energy. Firstly, the cosmological constant as a
candidate for dark energy will be excluded and dark energy must be
dynamical. Secondly, the simple dynamical dark energy models
considered vastly in the literature like the
quintessence\cite{pquint,quint} or the
phantom\cite{phantom,bigrip,Carroll:2003st,piao} can not be
satisfied either.

In the quintessence model, the energy density and the pressure for
the quintessence field are
\begin{equation}
 \rho=\frac{1}{2}\dot Q^2+V(Q)~,~~p=\frac{1}{2}\dot Q^2-V(Q)~.
\end{equation} So, its equation of state $W=p/\rho$ is in the range $-1\leq
W\leq 1$ for $V(Q)>0$. However, for the phantom \cite{phantom}
which has the opposite sign of the kinetic term compared with the
quintessence in the Lagrangian (we use the convention
$(+,~-,~-,~-)$ for the sign of the metric), \begin{equation}
 \mathcal{L}=-\frac{1}{2}\partial_{\mu}Q\partial^{\mu}Q-V(Q)~,
 \end{equation}
the equation of state $W=(-\frac{1}{2}\dot
Q^2-V)/(-\frac{1}{2}\dot Q^2+V)$ is located in the range of $W\leq
-1$. Neither the quintessence nor the phantom alone can fulfill
the transition from $W>-1$ to $W<-1$ and vice
versa.\footnote{Although the k-essence\cite{k-essence} like models
can have $W<-1$\cite{MMOT04}, it has been proved later by
Ref.\cite{Vikman} to be difficult to get $W$ across -1 during
evolution.} But at least a system containing two fields, one being
the quintessence with the other being the phantom field, can do
this job. The combined effects will provide a scenario where at
early time the quintessence dominates with $W>-1$ and lately the
phantom dominates with $W$ less than $-1$, satisfying current
observations. As an example, we consider a model:
\begin{eqnarray}\label{double}
 \mathcal{L}& =& \frac{1}{2}\partial_{\mu}\phi_1\partial^{\mu}\phi_1-\frac{1}{2}\partial_{\mu}\phi_2\partial^{\mu}\phi_2
 \nonumber\\
 & &-V_0[{\rm exp}(-{\lambda\over m_p}\phi_1)+{\rm exp}(-{\lambda\over
 m_p}\phi_2)]~,
 \end{eqnarray}
where $\phi_1$ and $\phi_2$ stand for the quintessence and
phantom. In Fig. \ref{SausageFig}, we illustrate the evolution of
the effective equation of state of such a system with $-\ln
(1+z)$.

\begin{figure}[htbp]
\begin{center}
\includegraphics[scale=0.45]{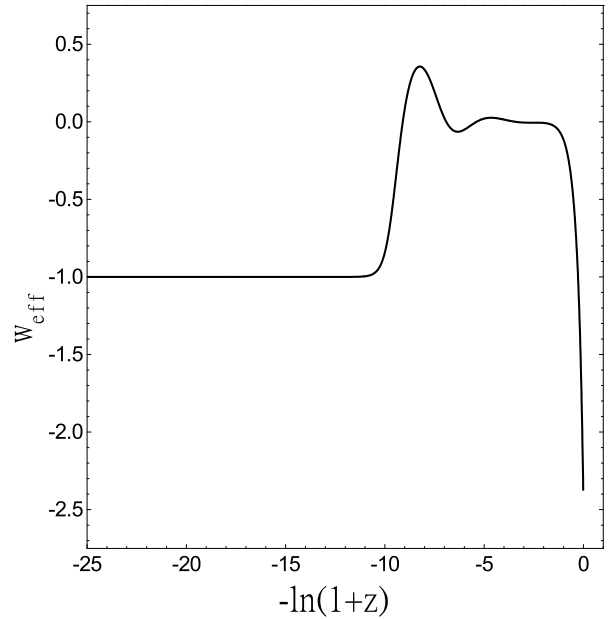}
\caption{ The evolution of the effective equation of state of the
double scalar fields given in Eq. (\ref{double}). The parameters
are chosen as: $V_0=8.38\times 10^{-126}m_p^4$, $\lambda=20$. We
set the initial conditions as: $\phi_{1i}=-1.7 m_p$,
$\phi_{2i}=-0.2292 m_p$, which lead to $\Omega_{m0}=0.30$,
$w_{{\rm eff} 0}=-2.44$. \label{fig:fig6}}
\end{center}
\end{figure}

In general to realize the transition of $W$ around $-1$, one needs
to consider models of dark energy with more complicated dynamics
and interactions with gravity and matter. This class model of dark
energy, which we dub ``Quintom", is different from the
quintessence or phantom in the determination of the evolution and
fate of the universe. Generically speaking, the phantom model has
to be more fine tuned in the early epochs to serve as dark energy
today, since its energy density increases with expansion of the
universe. Meanwhile the Quintom model as illustrated in Fig.5 can
preserve the tracking behavior of
quintessence\cite{Wangtrk,scaling}, where less fine tuning is
needed. We will leave the detailed investigation of the Quintom
models in a separated publication \cite{prepare1}, however will
mention briefly two of the possibilities below in addition to the
one in Eq. (9). One will be the scalar field models with
non-minimal coupling to the gravity \cite{superquin,prepare2}
where the effective equation of the state can be arranged to
change from above -1 to below -1 and vice versa. For a single
scalar field coupled with gravity minimally, one may consider a
model with a non-canonical kinetic term with the following
effective Lagrangian \cite{prepare1}:
 \begin{equation}
\mathcal{L}=\frac{1}{2}f(T)\partial_{\mu}Q\partial^{\mu}Q-V(Q)~,
 \end{equation}
where $f(T)$ in the front of the kinetic term is a dimensionless
function of the temperature or some other scalar fields. During
the evolution of the universe when $f(T)$ changes sign from
positive to negative it gives rise to an realization of the
interchanges between the quintessence and the phantom scenarios.

\begin{figure}[htbp]
\begin{center}
\includegraphics[scale=0.45]{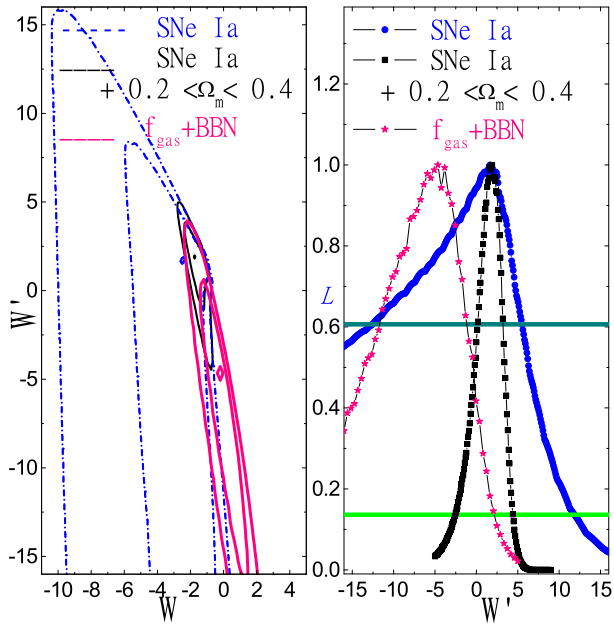}
\caption{ $2 -$ dimensional and $1 -$ dimensional constraints on
$W'$ in light of the data $f_{gas}$ from Chandra observations from
Ref. \onlinecite{Allen04}, with SNe Ia constraints also shown for
comparison. \label{fig:fig6}}
\end{center}
\end{figure}

Recently Allen {\it et al.} \cite{Allen04} have provided new
observational data basing on Chandra measurements of the X-ray gas
mass fraction in 26 X-ray luminous  galaxy clusters.  Under the
assumption that the X-ray gas mass fraction measured within
$r_{2500}$ is constant with redshift  the  $f_{gas}$ data  in the
range $0.07<z<0.9$ can be used directly to constrain cosmological
models. We use their data and fit to Model A, we set the same
gaussian prior on $\Omega_{baryon}$ as Ref.~\onlinecite{Allen04}
meanwhile varying h uniformly in range $0.51\sim 0.93$ and
$0<\Omega_m<1$. The $\chi^2$ value is defined as \begin{eqnarray}
\chi^2& =& \left( \sum_{i=1}^{26} \frac{\left[f_{\rm gas}^{\rm
SCDM}(z_{\rm i})- f_{\rm gas,\,i} \right]^2}{\sigma_{f_{\rm
gas,\,i}}^2}\nonumber \right)\end{eqnarray}
\begin{eqnarray}
+\left(\frac{\Omega_{\rm b}h^2-0.0214}{0.0020} \right)^2 +
\left(\frac{b-0.824} {0.089} \right)^2.
\end{eqnarray}
\vspace{0.1cm} For a detailed description of the data and fitting
see Ref.~\cite{Allen04}. We delineate the resulting $2 -$
dimensional and $1 -$ dimensional plots in Fig.6, together with
previous SNe Ia results for comparison (here we do not include age
constraint for simplicity). Very interestingly we get $W' < 0$
with the center value $(W,W')=(-0.2,-4.875)$ when varying fine
grids, which indicates the universe may not be accelerating today
but in the very near past. However the smallest redshift of
$f_{gas}$ data gives $z=0.077900$\cite{Allen04}, this shows also
some poor parameterization of model A. As shown in our paper and
also vastly in the literature the results of fitting depends
somewhat in the parameterizations\cite{Linde04}. However, in any
case if nonzero $W'$ gets more favored with the accumulation of
observational data, it gives strong implications for dark energy
"metamorphosis"\cite{sahni03}. We also find although the two data
sets give consistent results to model A, there seems to be some
discrepancy. In Fig.6, the contours do overlap in 1$\sigma$
regions, but only in a small area and the right panel is more
distinctive: $f_{gas}$ favors a negative $W'$ at more than
1.3$\sigma$ while a positive $W'$ is favored at around 1.1$\sigma$
with the prior on $\Omega_m$ for SNe Ia. Basically the X-ray data
probe the late behavior of the angular-diameter distance and SNe
Ia probes the luminosity distance. There seems to be some
discrepancy between them and this may possibly be due to some new
physics\cite{CKT02,BK03,FLLZ04}.

 In summary in this paper we consider the effect of cosmic
age and supernova limits on the variation of $W$. Our results show
that age limit plays a significant role in lowering the variation
of amplitude on the equation of state. Current SNe Ia observation
seems to favor a variation of $W$ from $>-1$ in the recent past
and $<-1$ today. If such a result holds on with the accumulation
of observational data, this would be a great challenge to current
cosmology. We give a simplest example of Quintom which can satisfy
the current implications on the equation of state on dark energy,
and discuss briefly the possibility of building Quintom models.

{\bf Acknowledgements:} We are indebted to Yun Wang for
enlightening discussions on supernova and age. We thank Tirth Roy
Choudhury, Zuhui Fan and Dragan Huterer  for helpful discussions
on supernova. We are grateful to  Steve Allen, Gang Chen, Ruth
Daly, Eric Linder,  Xingchang Song, Alexei Starobinsky, Jun'ichi
Yokoyama and Zong-Hong Zhu for discussions. We thank Mingzhe Li
for hospitable help and Xiaojun Bi, Hong Li and Yunsong Piao for
useful discussions. This work is supported in part by National
Natural Science Foundation of China under Grant Nos. 90303004 and
19925523 and by Ministry of Science and Technology of China under
Grant No. NKBRSF G19990754.

\end{document}